# Preparing for the Next Pandemic: Investigation of Correlation Between Food Prices and the COVID-19 Pandemic from Global and Local Perspectives

Yufei Zhao, Chao Huang, Jiebo Luo

## Abstract

The coronavirus disease (Covid-19) has caused enormous disruptions to not only the United States, but also the global economy. Due to the pandemic, issues in the supply chain and concerns about food shortage drove up the food prices. According to the U.S. Bureau of Labor Statistics, the prices for food increased 4.1% and 3.7% over the year ended in August 2020 and August 2021, respectively, while the amount of annual increase in the food prices prior to the Covid-19 pandemic is less than 2.0%. Previous studies show that such kinds of exogenous disasters, including the 2011 Tohoku Earthquake, 9/11 terrorist attacks, and major infectious diseases, and the resulted unusual food prices often led to subsequent changes in people's consumption behaviors. We hypothesize that the Covid-19 pandemic causes food price changes, and the price changes alter people's grocery shopping behaviors as well. To thoroughly explore this, we formulate our analysis from two different perspectives, by collecting data both globally, from China, Japan, the United Kingdom, and the United States, and locally, from different groups of people inside the U.S. Particularly, we analyze the trends between food prices and Covid-19 as well as between food prices and spending, aiming to find out their correlations and the lessons for preparing the next pandemic.

## Introduction

In the past two years, the Covid-19 pandemic has severely impacted people's lives, in the ways of how people eat, travel, work, and socialize. As a necessity to living, the price of food fluctuated more drastically than before, which would gradually change people's shopping behaviors and thus the spending on food. For example, rising beef prices may force people to choose other alternatives, such as pork and chicken. Therefore, it is interesting and useful to investigate the correlation between food price changes and the trend of the Covid-19 pandemic.

To this end, we will take a look at the price changes of different food categories, which can be broadly organized into specific subcategories, including meat, fruits, vegetables, dairy products, oil, cereals, sweets, and seafood. There are many reported factors that may impact consumer expenditure on food at home (FAH) and food away from home (FAFH), e.g., income (Lusk 2019), region, age (Ziliak 2021), neighborhood severity level measured by the number of local Covid-19 cases and deaths (Baker et al. 2020; Huang, Sant'Anna, and Etienne 2021), and household size and composition (Statista 2020). These factors all account for different expenditure behaviors. Among those, we choose income, region, number of children, and household size for further analysis in our study, since the relationship between the Covid-19 severity level and food prices, food expenditure, and spending behaviors are already well studied. In the *local* perspective, we group consumers based on the above four attributes (income, region, number of children, and household size) and compare expenditure behaviors of different consumers before and after the Covid-19 pandemic.

Since the Covid-19 pandemic is a global rather than local event, the food price-Covid-19 relationship may vary in different counties, states, and even countries. In our study, we perform *global* analysis of the relationship. Apart from the United States, we collect data from the Unites Kingdom, China, and Japan, which have distinctively different dietary habits and different policy responses towards Covid-19. We intend to examine the food price changes under different political policies and security levels throughout the Covid-19 pandemic and identify the food price change trends and patterns in these countries.

To summarize, in this study, we intend to answer the following questions:

- How does the food price of each food subcategories change before and during the Covid-19 pandemic?
- How do consumers with different attributes change their food shopping behaviors?
- What are the differences and similarities of the changes in food prices in countries with different political policies and security levels?

## Related Work

As described in the introduction, the Covid-19 pandemic has drastically affected people's lives. Recent works on food supply chains suggested that some problems caused by Covid-19, such as labor shortage (Richards and Rickard 2020; Thilmany et al. 2021) and reduced shipments (Pena-Levano et al. 2020) have led to higher food prices. As a result, people's lifestyles suddenly changed. A series of studies on shopping habits (Mead et al. 2020; Melo 2020; Knotek II et al. 2020; Ellison et al. 2021) also show that Covid-19 also change the way of buying, as consumers tend to stock up on groceries during the pandemic.

To further figure out the relationship between food prices and consumers' spending behaviors, we conduct studies on consumers' spending from different household groups. Prior works show that sudden exogenous events often resulted in temporal shopping behavior changes, including the 2003 SARS pandemic (Yang et al. 2020), the 2011 Tohoku Earthquake (Hori and Iwamoto 2014), the 9/11 terrorist attacks (Dube and Black 2010), and the Fukushima Nuclear Accident (Frank and Schvaneveldt 2014). The lesson learned from these events is that the spending pattern changes depending on multiple factors, including consumer income, consumer experience, education levels, etc. Thus, how the food price change affects consumers' spending behavior during the Covid-19 pandemic still needs to be explored. A most recent work (Huang, Sant'Anna, and Etienne 2021) shows an analysis on household purchasing trends after the economy had partially reopened. Compared to this work, we not only study how consumers' spending behavior changes on food, but also analyze the correlation between food price and spending pattern changes.

## Datasets

We employ Consumer Price Index (CPI) to investigate the food price changes, so that we can compare price changes among different countries, which have different units of measurement. By definition, the Consumer Price Index (U.S. Bureau of Labor Statistics 2021) is a measure of the average change over time in the prices paid by consumers for a market basket of consumer goods and services, such as transportation, food, and medical care. The basic formula is shown below:

$$CPI_t = \frac{C_t}{C_0} \times 100 \qquad (1)$$

where: $CPI_t$ is the Consumer Price Index in the current period; $C_t$ is the cost of the market basket in the current period; $C_0$ is the cost of the market basket in the base period.

CPI is one of the most frequently used statistics for dealing with price change problems. There are many ways to use CPI for price change. Take the publications and reports released by the U.S. Bureau of Labor Statistics as an example. They have unadjusted 12-month percent change of CPI, unadjusted 1-month percent change in CPI, seasonally adjusted 12-month percent change in CPI, and seasonally adjusted 1-month percent change in CPI. We use an unadjusted 1-month percent change in CPI to measure the price change in our study.

We combine and use data from the following five websites. We collect food CPI of the United States, the United Kingdom, China, and Japan from January 2019 to September 2021, and U.S. food expenditure from April 2020 to June 2021. We then consolidate data of CPI of four different countries into the unified formats and transform the weekly U.S. expenditures into monthly U.S. expenditures.
- U.S. Bureau of Labor Statistics: https://www.bls.gov/.
  We get CPI for food subcategories in the United States from this website. The website provides CPI and food prices for overall food category, common food items, and food subcategories.
- U.S. Census Bureau: https://www.census.gov/en.html.
  We get average weekly food spending of United States consumers with different income, household size, number of children, and location from this website.

- National Bureau of Statistics of China: http://www.stats.gov.cn/english/.
  We get CPI for food subcategories in China from this website. The website provides CPI for common food items, food categories, and other commodities (clothing, residence, etc.).
- Office for National Statistics: https://www.ons.gov.uk/.
  We get CPI for food subcategories in the United Kingdom from this website. The website provides CPI for common food items and food categories.
- e-State/Statistics Bureau of Japan: https://www.stat.go.jp/english/.
  We get CPI for food subcategories in Japan from this website. The website provides CPI for common food items, food categories, and other commodities (clothing, residence, etc.).

# Methodology

## Pearson product-moment correlation

For the global perspective part, we want to capture the relationship between each pair of different food sub-categories and their importance in the overall food category for different countries. The formula (Glen 2021) is the following:

$$\rho_{xy} = \frac{Cov_{xy}}{\sigma_x \sigma_y} \qquad (2)$$

where $\rho_{xy}$ is the Pearson product-moment correlation coefficient, and $Cov(x, y)$ is the co-variance of two food sub-categories $x$ and $y$, $\rho_x$ and $\rho_y$ are the standard deviation of x and y.

With a higher Pearson product coefficient, we claim that the two food sub-categories have a stronger association strength. With a higher Pearson product coefficient with the overall food category, we claim that the food sub-category has a stronger association strength or greater importance in the food category.

## Trend and Differential Stationary

In general, it is necessary for time series data to be stationary to satisfy the assumption of time series analysis models. A stationary time series is a time series whose properties do not depend on the time at which the series is observed. Time series with trend and seasonality are not stationary. In our case, there is no seasonality. Our Covid-19 data and food price data during Covid-19 do not experience regular and predictable changes over year for every calendar year, because the Covid-19 pandemic is an unpredictable and temporal disease in the recent two years. However, we need to consider the trend. There are some common ways to remove trends from time-series data, such as log transformation, power transformation, and difference transformation. We apply the difference transformation (Hyndman and Athanasopoulos 2018), which works by computing the differences between consecutive observations. The formula is given by:

$$y_t' = y_t - y_{t-1} \qquad (3)$$

The differentiated series will have only $t - 1$ values, since it is not possible to calculate a difference for the first observation. For the countries whose time series still do not appear to be stationary after the first differentiation, we perform a secondary-order differentiation and a general de-trend function to obtain a nearly stationary time series. The formulas for the secondary-order differentiation and de-trend function are shown below:

secondary-order differentiation:
$$y_t'' = y_t' - y_{t-1}' = y_t - 2 \times y_t - 1 + y_{t-2} \qquad (4)$$
and the de-trend function:
$$y = detrend(x) \qquad (5)$$

In this case, the time series will have $t - 2$ values.

The ADF test (Dickey and Fuller 1979) and KPSS test (Kwiatkowski et al. 1992) are the most commonly used statistical tests to analyze and check the stationary of a series. They make hypotheses about data and inform the degree to which a null hypothesis can be rejected or fail to be rejected.

The ADF test is conducted with the following hypotheses:
- Null Hypothesis (H0): Series is non-stationary.
- Alternate Hypothesis (HA): Series is stationary.

If test statistic < critical value or $p$-value < 0.05, we reject the Null Hypothesis (H0) and state that the time series is stationary. If the null hypothesis is failed to be rejected, this test may provide evidence that the series is non-stationary.

The KPSS test is conducted with the following hypotheses:
- Null Hypothesis (H0): Series is stationary.
- Alternate Hypothesis (HA): Series is non-stationary.

Note that the hypothesis is reversed in the KPSS test compared to the ADF Test. If test statistic < critical value or $p$-value < 0.05, we reject the Null Hypothesis (H0) and state that the time series is non-stationary. If the null hypothesis is failed to be rejected, this test may provide evidence that the series is stationary.

### Granger Causality

The Granger causality test (Granger 1969) is a statistical hypothesis test for determining whether one variable in the time series is useful for forecasting another in the multivariate time series with a particular lag. Our study chooses lag equals 1. A prerequisite for performing the Granger Causality test is that the input time series should be stationary (Chen et al. 2018). Granger Causality Test is conducted with the following assumptions: Null Hypothesis (H0): time series $x$ does not explain the variations in time series $y$. (Or we say, $x_t$ does not Granger-cause $y_t$.) Alternate Hypothesis (HA): time series $x$ explains the variations in time series $y$. (Or we say, $x_t$ Granger-causes $y_t$.) This test produces an F test statistic with a corresponding $p$-value. If the $p$-value < 0.05, we reject the Null Hypothesis (H0) and state that $x_t$ Granger-causes $y_t$. If the null hypothesis is failed to be rejected, this test may provide evidence that $x_t$ does not Granger-cause $y_t$.

## Experimental Results and Discussion

### Global Perspective

Our global perspective part contains analyses for China, Japan, the United Kingdom, and the United States and is carried out in two steps.

In the first step, the CPI datasets of the eight food subcategories (*exception: seven food sub-categories for China) and the Covid-19 statistics about confirmed cases and confirmed deaths of all four countries are overviewed and explored using the Pearson product-moment correlation. The results of the Pearson product-moment correlation are displayed with heatmaps in Figure 1.

From the resulted heatmaps (Figure 1), we can obtain information about the importance of each food sub-categories. Ordinarily, the food sub-categories with a higher positive correlation with the general food tend to be relatively more important. The reason is that these food sub-categories are more likely to have their prices moving in the same direction with the general food category, say increasing or decreasing together. In other words, when the price of these food sub-categories increases or decreases, the price of the general food category has a high probability to increase or decrease, respectively. For the assisted and less bought food, however, their price index does not affect the general food category a lot and thus perhaps does not this kind of synchronous price change. Hence, this strong association suggests a higher greater importance.

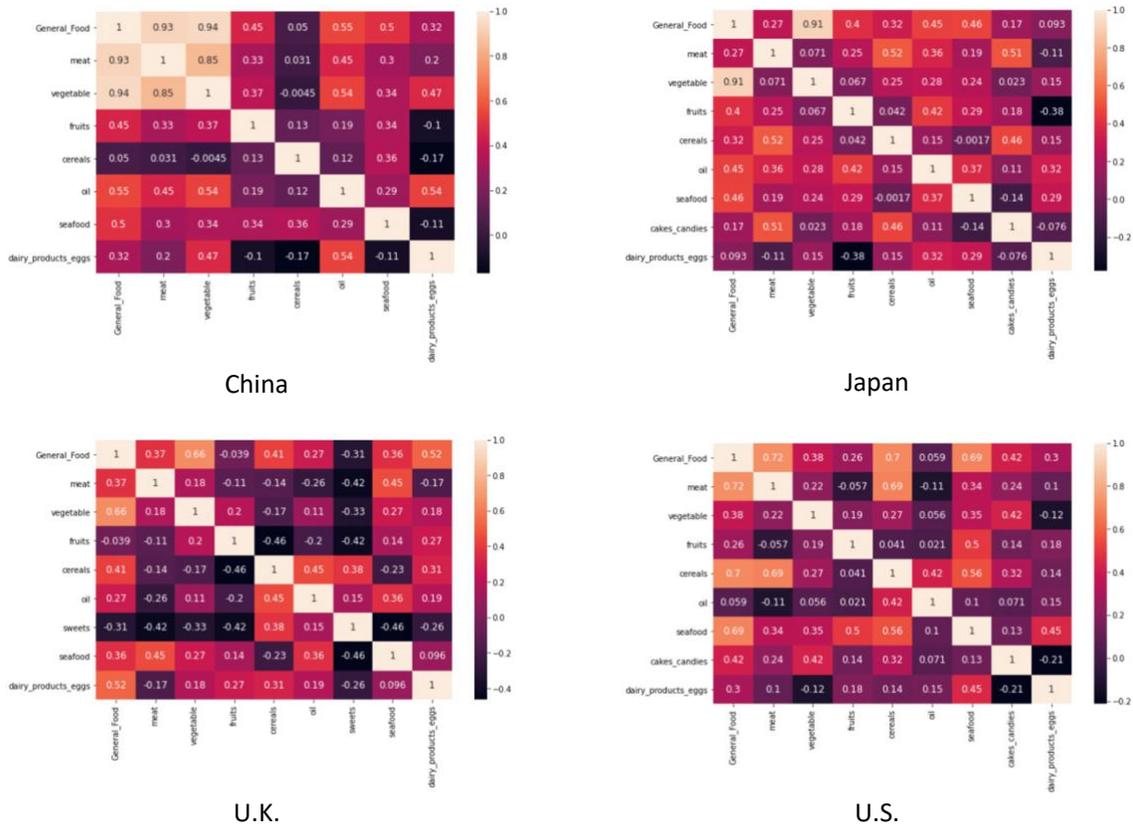

Figure 1: Food CPI heatmaps for the correlation between (#number) food sub-categories

For China, meat and vegetables are particularly important, with a correlation above 0.90, which is much higher than other food sub-category pairs in China. As shown in Figure 2, the price of meat and the price of vegetables after the outbreak of Covid-19 pandemic display a similar increasing trend and decreasing trend. As the price is heavily driven by the demand, we can infer that Chinese consumers often increase or decrease their demand for vegetables and meat together, and the vegetables and meat possibly are complementary food and both valuable in the Chinese diet. These high correlations are also rare in other countries. However, it may be a little surprising that the cereals, which include the Chinese staple food rice, have a relatively low correlation with the general food category. This may be explained by the stable price level of cereals during the Covid-19 pandemic in China as shown in Figure 2.

For Japan, vegetables take an important place in their diet, due to a 0.91 correlation with the general food category. The seafood, fruits, oil, and cereals are in the following places. As an irreplaceable traditional Japanese food, seafood (sushi) does not have a correlation above 0.50, perhaps due to the relatively stable price level similar to Chinese cereals (Figure 2). Surprisingly, we notice that the oil price often has a quick response to the Covid-19 confirmed cases. The local maximum points of the Japanese oil price trend all correspond to a remarkably large number of monthly confirmed cases compared to previous months, such as April 2020, January 2021, and May 2021. This may imply that oil is the good that Japanese demand most when the pandemic severity gets worse.

For the United Kingdom and the United States, there are no particular food sub-categories that have a very high correlation and strong association with the general food category. Relatively, vegetables and dairy products are important diet sources for United Kingdom citizens, while meat, cereals, and seafood are important for Americans. Generally, these results accord with our expectations. However, similar to the Japanese people's favor of oil, United Kingdom citizens drive up the price of dairy products when there is a new wave, but with a little slower response. The local maximum points of the price of dairy products often appear 1 month after a notable percentage change in monthly confirmed cases. United States citizens don't have a constant response to Covid waves in terms of buying specific food. The only noticeable response is the high demand for meat immediately after the outbreak of Covid-19 from April 2020 to June 2020.

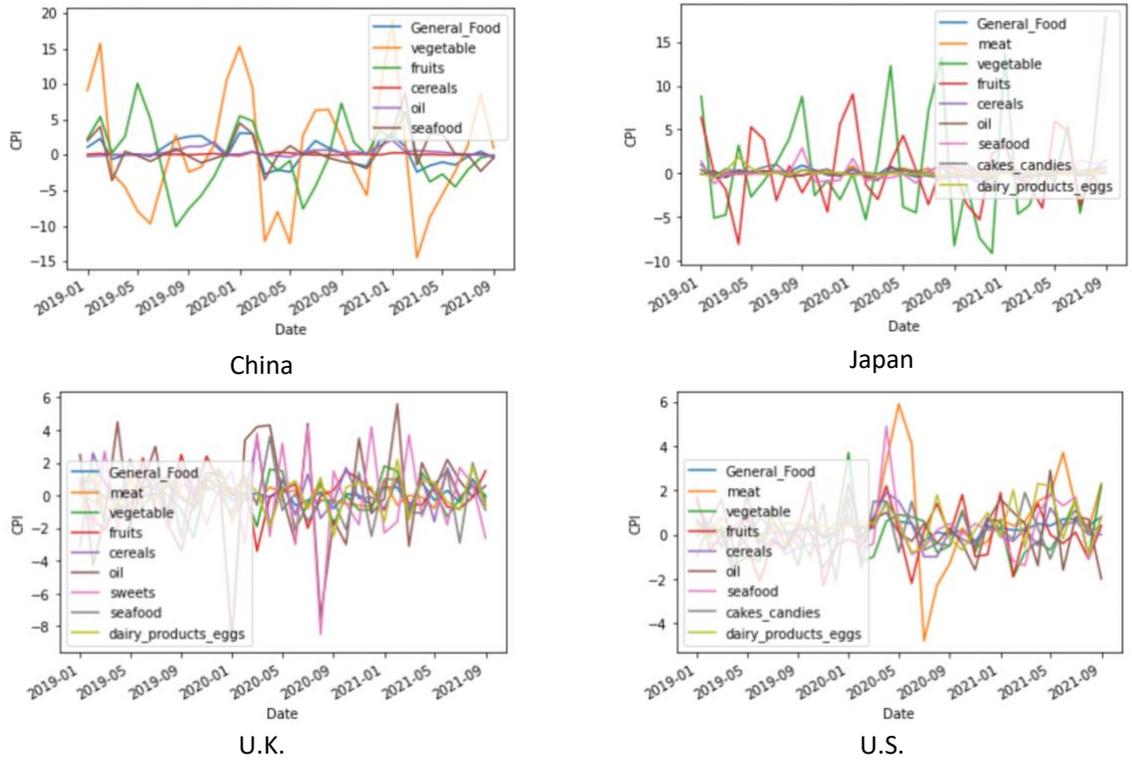

Figure 2: Price trends on different food sub-categories from Jan 2019 to Sept 2021

In the second step, we create time series for all the food-related variables and Covid-19-related variables in our datasets and perform the Granger Causality Test on the time series to determine whether the Covid-19 conditions predict and influence the food prices. According to the Methodology Section, we detrend the time series to make them stationary before the Granger Causality Test. The key variable we choose to represent the Covid-19 conditions and apply as a predictor is the monthly confirmed cases alone, ignoring the total confirmed cases, total confirmed deaths, and monthly confirmed deaths.

According to the Granger Causality Test, if the *p*-value is less than 0.05, we have strong evidence that the Covid19 condition influences that specific food price. The results of the Granger Causality Test for all four countries are shown in Table 1.

| country | food | vegetables | fruits | meat | cereals | dairy | seafood | cake | oil |
|---|---|---|---|---|---|---|---|---|---|
| China | **0.0027** | **0.0239** | **0.0005** | **0.0248** | **0.0105** | 0.4348 | **0.0000** | - | **0.0002** |
| Japan | **0.0015** | **0.0107** | 0.6827 | 0.4050 | 0.5074 | **0.0000** | *0.0618* | 0.4318 | **0.0048** |
| UK | **0.0001** | **0.0189** | 0.1033 | 0.4886 | 0.4764 | **0.0468** | *0.0764* | 0.2118 | 0.1122 |
| US | 0.4067 | 0.4905 | 0.2921 | 0.5434 | **0.0406** | 0.0148 | 0.3209 | 0.0012 | **0.0000** |

Table 1: *p*-values for China, Japan, the United Kingdom, and the United States (Strike-through denotes the time series is not stationary; dash denotes no number is for the column)

From the results, we can say that the Covid-19 conditions influence all the four countries. This result somewhat epitomizes the food industry conditions during the Covid-19 pandemic. The panic buying and hoarding behaviors have been seen globally, many supermarket shelves with key food such as rice, pasta, canned goods, and sanitation supplies are emptied (Donnelly 2020). Moreover, many restaurants, bars, and cafes have closed due to the global stay-at-home policies and reduced outside mobility; this brings even more pressure to the food supply chain (Hobbs 2020). The joint impacts on the demand chain and supply chain drive the food prices up. One important similarity among the results of the four countries is that the vegetable price is influenced by the Covid-19 conditions, while the price of meat, another vital and necessary food intake, is almost not influenced. This situation may be referable to the difference in production mode (Deconinck, Avery, and Jackson 2020). Vegetables are much more labor-intensive and season-dependent than meat, which means that the limitations on the mobility and stay-at-home policy reduced the availability of seasonal workers and disrupted the timeline for planting and harvesting vegetables, but not affected meat production that much. Furthermore, the cut-off of air transportation makes it difficult for seed and fertilizer supply, the decisive factors, for vegetable production. Unfortunately, for all the four countries, the experiment result also suggests that a change in the vegetable price would lead to a change in the fruits price, which is also validated by the synchronous but a little lagging increase or decrease of the fruits price trend compared to the vegetable price trend shown in Figure 2.

We can also see the differences in the degree to which the Covid-19 conditions influence food prices. The food prices in the China market are mostly struck, with almost all the food sub-categories being Granger-caused by the Covid-19 conditions. Part of the price changes is due to government suggestions. When the number of confirmed cases or deaths in some local areas, the central government will urge the local authorities to encourage families to stockpile food and other daily necessary goods. This authority encouragement even induces panic buying and intensifies the food price fluctuations with Covid-19 conditions (McNamara 2021). The perception of threats and anxiety about uncertainty also play an important role in the China market (Roy and Chakraborty 2021). Besides, China's strict quarantine policy worsens the food supply chain. Many truck drivers are reluctant to make deliveries to virus-hit areas, in order to avoid the 7 days or even longer quarantine (Kawate 2022). The food price rises and fluctuations in Japan were just behind China, which is evidenced by the CPI axis. Though with relatively less strict quarantine and travel restrictions compared to China, Japan employed the state of emergencies several times in 2020 and 2021 (Otsuka 2021), which greatly hindered transportation. China's strict Covid-19 lockdowns may also be a partial reason for Japanese soaring raw material costs and supply chain disruptions. On the contrary, the Covid-19 conditions in the United States do not have a clear impact on food prices. Large price movements, such as the meat price inflation, appeared between April and September 2020. This early period price shocks then returned to stable levels. The situation may be explained by US's role in the global food trade. In order to control the spread of the Covid-19 pandemic, both air and ground transportation becomes inconvenient and costly. Therefore, the countries that previously relied on food imports to some extent will get into trouble and experience some food shortages, including China, Japan, and the United Kingdom, while the countries that primarily export food (Oteros-Rozas et al. 2019) meet fewer food troubles. In addition, there are fewer rules regarding traveling and vaccination in the United States. The local transportation among U.S. states is less hindered compared to that of China and Japan. The explanation applies to the United Kingdom as well.

Because of the different price movements and responses to Covid-19 in the four countries, we may be able to summarize the strategies for regulating the food market before and during a pandemic. Rather than "fighting against the rising food prices", we should "avoid food shortages". A stable supply chain is most important. Vegetables are one of the most vulnerable and highest in demand categories during the pandemic, and their price is greatly influenced by Covid-19. Since the countries often meet the waves at different times, it may be helpful for the nearby countries to trade for vegetables to overcome the shortage of vegetables. Also, the price of fruits may change right away after the change in the price of vegetables. Therefore, as the government notices a large price change in vegetables, they may need to immediately prepare some fruits for storage. In particular, for countries with similar eating habits to Japan, their governments may need to trade for oil, if they somehow predict a new wave or hear that a new wave comes to some countries. Countries like the United Kingdom need to prepare dairy products and eggs for storage after the Covid wave. Furthermore, from the

China and Japan cases, we infer that the frequent and nonstop enforcement of "stay-at-home" and strict travel requirements increase people's concerns of uncertainty and danger, drive up the scale of food stockpiling, and harm the food supply. Therefore, the policymakers should consider the public mood and adjust their policies and food supply accordingly in a timely fashion.

**Local Perspective**

In this section, we divide the analysis of the correlation between food prices and *local* consumer spending patterns into two parts: 1) the food CPI change before and during the pandemic across different regions in the US; 2) the correlation between food price changes and certain US household, that is, whether food price fluctuation has affected the spending on food for the given household type.

**Food CPI changes in different census divisions.**

We collect the food CPI data of different regions in the US from Jan 2018 to Oct 2021. Given the data, we can show how the food price changed before and during the Covid-19 pandemic, indicating the impact that Covid-19 had on food prices.

The data is collected from different Census Bureau regions (Wikipedia 2020), including {*New England, MidAtlantic, East North Central, West North Central, South Atlantic, East South Central, West South Central, Mountain, Pacific*}. Note that these census regions are widely used for data collection and analysis.

In Figure 3, we select four different timestamps before and during the pandemic, namely *Jan 2018*, *March 2019*, *March 2020* and *March 2021*, to provide an overview of how the food prices change in different stages. For the range of CPI change, we normalize it to [-1, 1], using color heat to represent the magnitude of change. Before the pandemic, the food CPI changes were slight Figure 3 (a,b). However, due to the outbreak of Covid-19, the CPI changed drastically (Figure 3 (c)). In Figure 3 (d), we observe a similar change pattern as before the pandemic, partially caused by the reopening and people's accustomed attitude towards Covid-19.

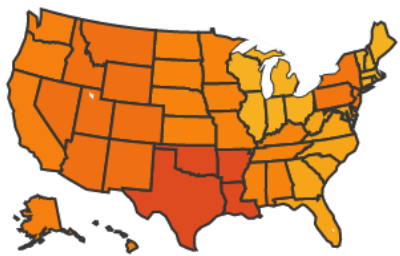

(a) Food CPI change in Jan 2018

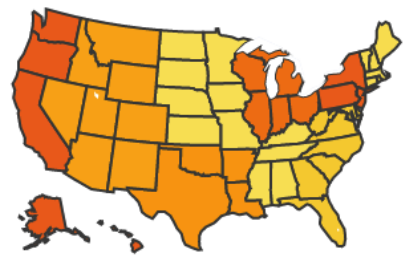

(b) Food CPI change in March 2019

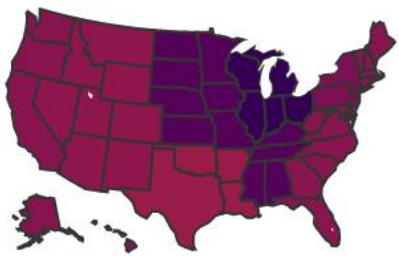

(c) Food CPI change in March 2020

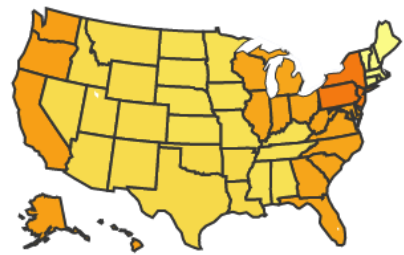

(d) Food CPI change in March 2021

Figure 3: Overview of Food CPI changes in different stages. For the range of CPI changes, we normalize it to [-1, 1], using color heat to represent the magnitude of change.

**Visualization of food spending trends of different households.**
Instead of only looking at the food price change, it is more interesting to explore the correlation between people's spending behavior and food prices. The data collected from U.S. Census Bureau provides us with four different angles to investigate the problem, that is, *household size*, *number of children*, *income*, and *region*. Note that, although the data we use do not contain more specific subcategories like the age of people in the household with different sizes and whether there are people in retirement or under 18, it still can provide a meaningful study on how different groups of people are affected by the Covid-19 pandemic. In the following, we will analyze the correlation between the change in food-at-home spending and the given household types.

- Household size
  We study the monthly spending on food at home for households of different sizes, starting from 2 people in the household. In Figure 4, we can find that all the candidates are affected by the Covid-19 pandemic, as they exhibit similar change points as Covid19. Since the blue curves represent spending on food at home, the similar pattern as Covid-19 indicates that when the monthly newly confirmed Covid-19 cases increase, people tend to stay at home and avoid eating out at food-away-from-home establishments, such as restaurants, school cafeterias, sports venues, and other eating-out establishments (USDA 2021). Thus, food at home spending goes higher along with the newly confirmed cases. Interestingly, it is shown that with more persons inside the household, the spending trend is smoother before the first change point. The previous survey on household size and the demand for food (Deaton and Paxson 1998) suggested that the larger households are better off at food control since they have the option of decomposing themselves into smaller units. There will also be substitution effects toward the shared foods, which are effectively cheaper for members of the larger household. In this case, even if the food price changed at the beginning of the pandemic, larger households can still control their spending on food. The larger the household is, the better they can do in sharing food, which corresponds to the observation in Figure 4.

- Number of children
  Similarly, we study the monthly spending on food at home for households with different numbers of children. Most of the spending patterns are similar to the trend of Covid-19. In other words, the spending on food at home went higher when the case of Covid-19 continuously increased. However, the two types of households performed differently. For households with four or more children, the spending on food at home increased at the beginning of the outbreak. There are generally two reasons: 1) Most children attended school or received care at home, increasing the need for childcare and home cooking; 2) Eating at home was perceived as safer especially for children, increasing the food at home spending (Ferrante et al. 2021).

- Income
  In Figure 6, the spending behavior of households with different levels of income is similar. As previous studies (French, Wall, and Mitchell 2010) showed, higher-income households spent significantly more money on both at-home and away-from-home food compared with lower-income households, especially on fruits/vegetables. In Figure 6, however, the curves denote the percentage change. Thus, the spending patterns between high- and low-income households still need to be explored.

- Regions
  In Figure 7, we visualize the at-home food spending trend of different regions, including *Northeast*, *Midwest*, *South*, and *West*. Interestingly, the spending patterns are generally categorized into two groups: {Northeast, South} and {Midwest, West}. The change in the former group is sharper than in the second one, which shows the difference between policies across different regions.

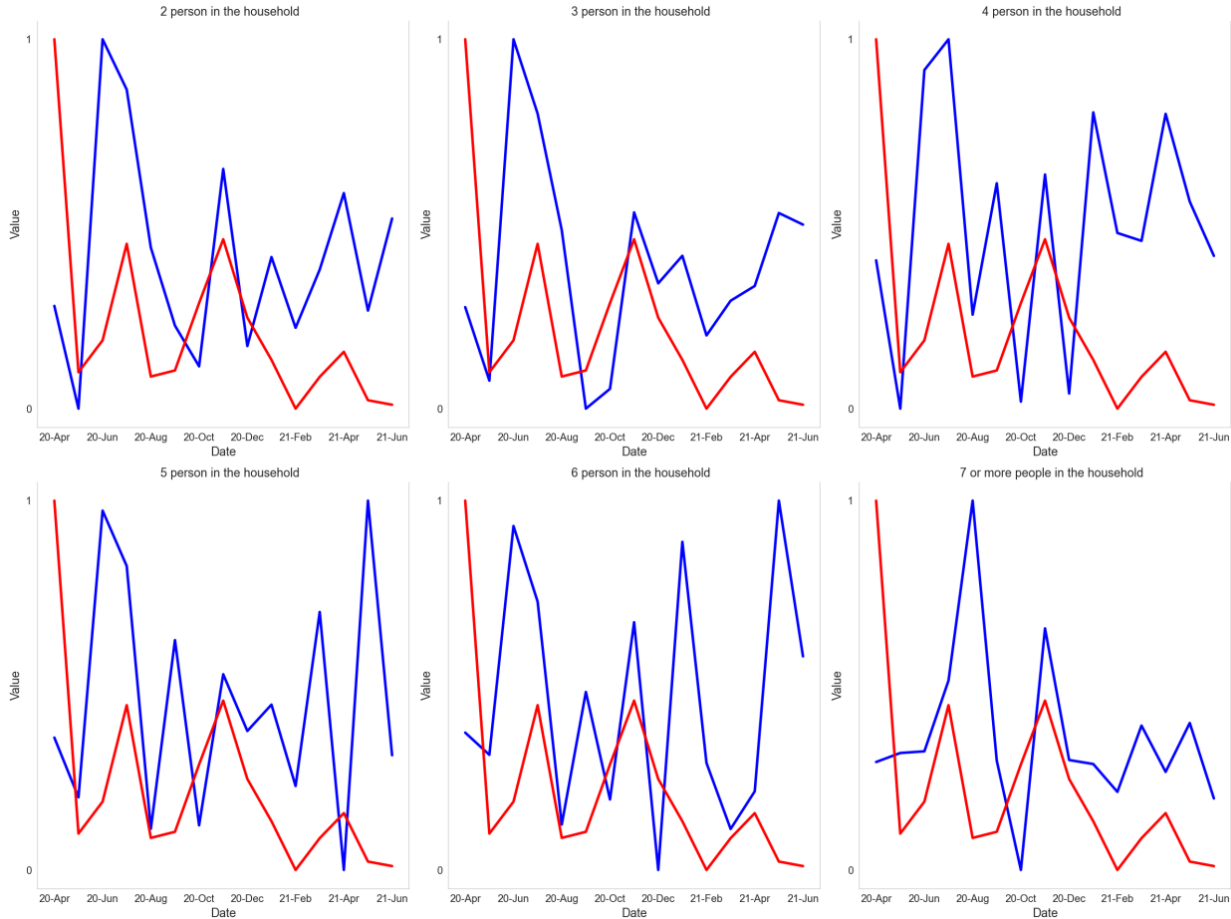

Figure 4: Curve of food spending for different household sizes (blue) and the COVID-19 trend (red)

| household size | 1 person | 2 people | 3 people | 4 people | 5 people | 6 people | 7 or more people |
|---|---|---|---|---|---|---|---|
| $p$-value | 0.3067 | 0.0767 | 0.1566 | **0.0022** | **0.0017** | **0.0034** | **0.0007** |

Table 2: $p$-values for household size

| number of children | No children | 1 child | 2 children | 3 children | 4 children | 5 or more children |
|---|---|---|---|---|---|---|
| $p$-value | **0.0000** | **0.0000** | 0.0653 | ~~0.0069~~ | ~~0.7414~~ | **0.0282** |

Table 3: $p$-values for households with different numbers of children (Strike-through denotes the time series is not stationary)

| income | Less than $25,000 | $25,000 to 34,999 | $35,000 to 49,999 | $50,000 to 74,999 |
|---|---|---|---|---|
| $p$-value | **0.0006** | **0.0085** | **0.0004** | **0.0069** |
| income | Less than $75,000 to 99,999 | $100,000 to 149,999 | $150,000 to 199,999 | $200,000 and above |
| $p$-value | **0.0000** | **0.0001** | 0.3482 | ~~0.1555~~ |

Table 4: $p$-values for households with different levels of income (Strike-through denotes the time series is not stationary)

| region | Northeast | South | Midwest | West |
|---|---|---|---|---|
| *p*-value | 0.0078 | 0.010 | 0.000 | 0.0000 |

Table 5: *p*-values for households from different regions

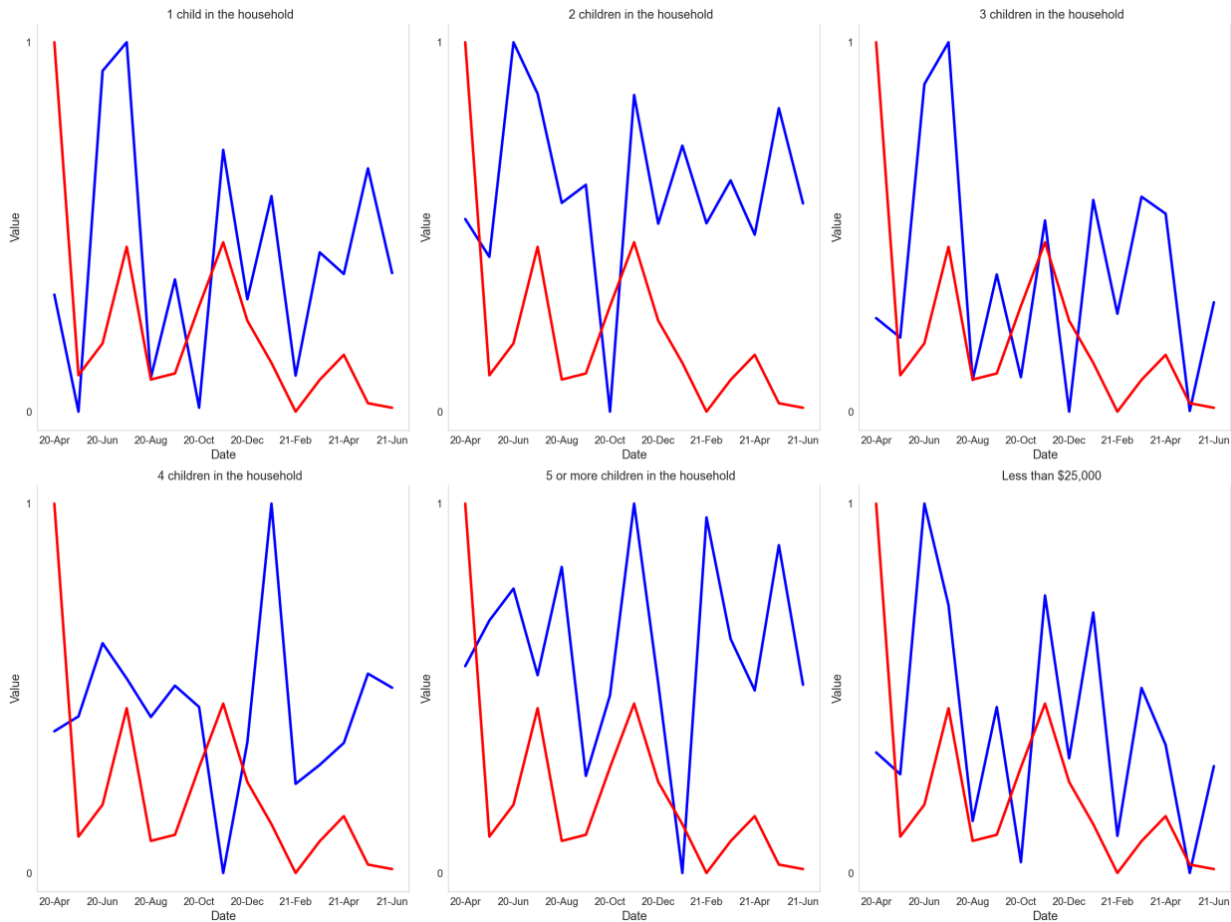

Figure 5: Curve of food spending for different numbers of children in the household (blue) and the COVID-19 trend (red)

**Statistical analysis between the food spending of different households and the food prices.**
In addition to the qualitative visualization, we also perform a statistical time series analysis for household spending and food price changes. Specifically, we follow the previous section and use the Granger Causality Test on the time series to determine whether the food price changes had impacted the food at home spending of different households. According to the Methodology Section, we differentiate the time series to make them stationary before the Granger Causality Test. According to the Granger Causality Test, if the *p*-value is less than 0.05, we have strong evidence that the food price change influences the spending of that specific household. The results for the Granger Causality Test for all the households are shown in Table 2, 3, 4, and 5.

In Table 2, we found that the food at home spending of households with 4, 5, 6, 7 or more people are more likely to be affected by the food price changes. We hypothesize the reason behind this phenomenon is that the consumer unit with less than four people is usually couples or single, and they are more like to order food away from home. Thus, their spending on food at home for them is less affected by the Covid-19 pandemic and the food price change.

In Table 3, we compute the correlation between households with different numbers of children and food prices and found that the spending of households with 2 children is less likely to be affected by the food price changes, as the p-value > 0.05.

*P*-value shown in Table 4 illustrates that higher-income households, e.g., with annual income from $150,000 to 199,999, have a lower correlation with the food price changes. A hypothesis is that higher-income households tend to spend more on both at-home and away-from-home food, especially on fruits and vegetables. However, due to the pandemic, there are reduced shipments (Pena-Levano et al. 2020) in the US, causing unstable fruit and vegetable prices in different regions since they need careful storage and shipment. In this case, the correlation between food spending of this particular group and the food price changes is harder to analyze. It needs to take a closer look at other factors that may jointly affect food price and spending, e.g., regions. Results in Table 5 indicate that the food price change influences the spending patterns for households in all of the four regions. Specifically, {Northeast, South} group has a much higher *p*-value of around 0.01 than the {Midwest, West} group, which is close to 0. This suggests that people in the Midwest and West regions shop similarly, while people in the Northeast and South regions tend to share similar shopping habits.

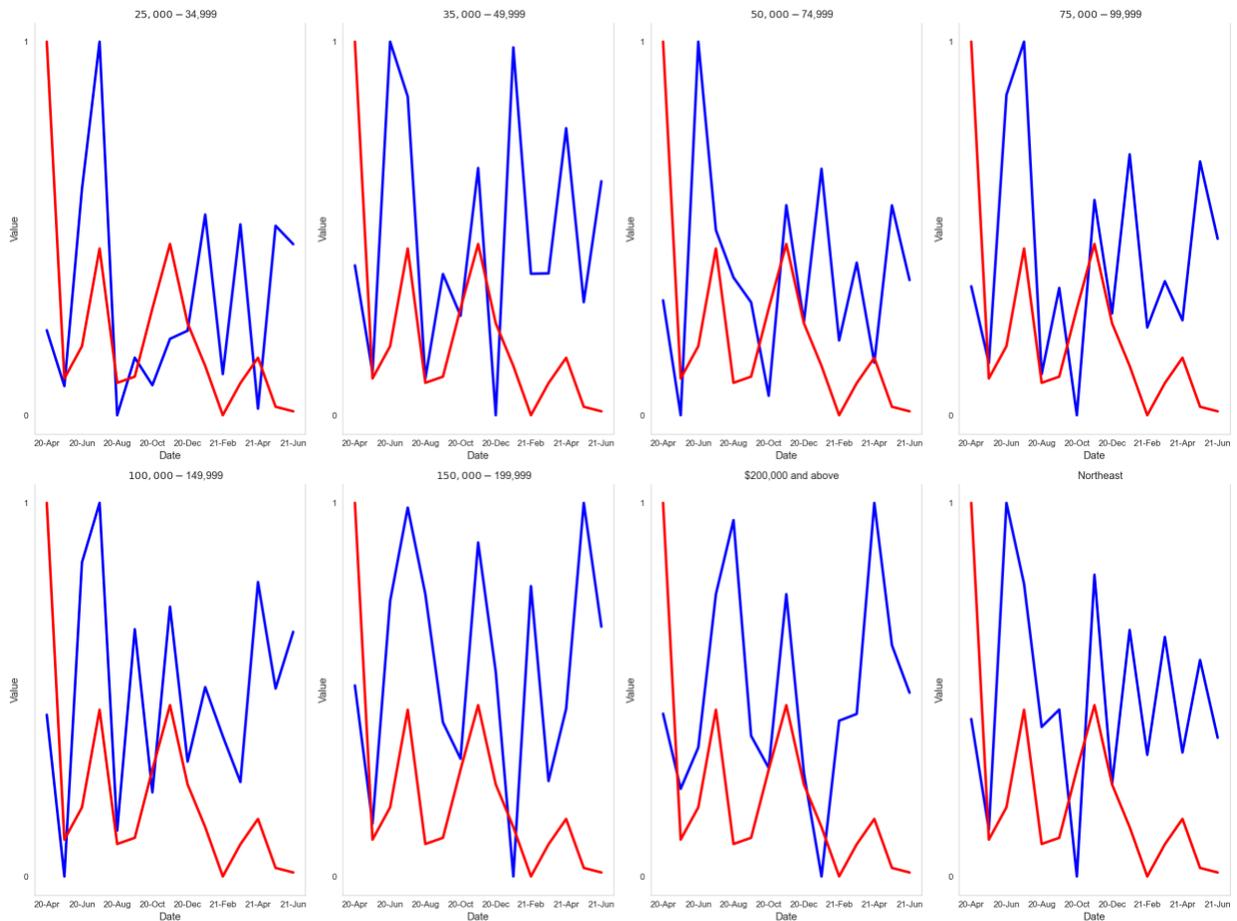

Figure 6: Curve of food spending for households with different incomes (blue) and the COVID-19 trend (red)

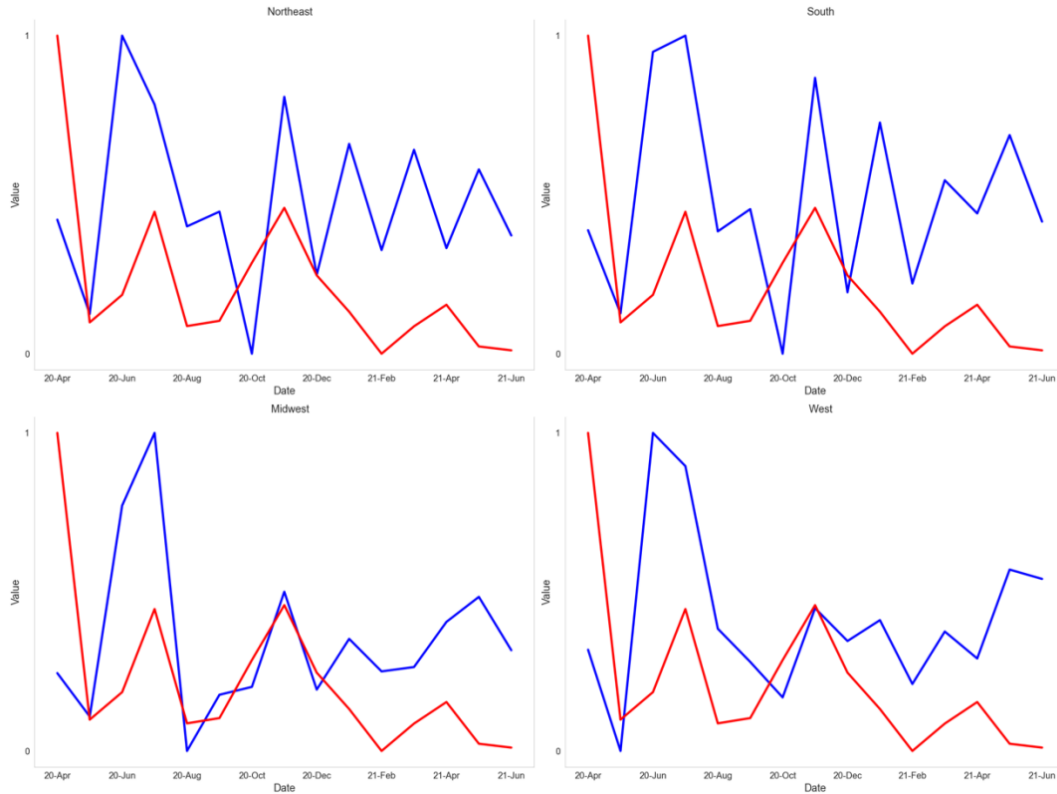

Figure 7: Curve of food spending for different regions (blue) and the COVID-19 trend (red)

## Conclusions

In this study, we have found that the Covid-19 pandemic severely affects people's lives, subsequently changing people's shopping behaviors. By analyzing the data from both local and global perspectives, we draw several important conclusions: 1) most countries saw their food prices influenced by the Covid-19 pandemic; 2) *vegetables are the most vulnerable food sub-category to the pandemic* and serve as one of the most important food intakes during the Covid- 19 pandemic, while another important food sub-category of meat, *contrary to the common complaints*, is one of the most resistant foods to the pandemic; 3) the spending on at-home food for *larger households* is more likely to be affected by the food price changes; 4) higher-income households exhibit different shopping habits from the lower-income households, showing that they are less likely to be affected by the Covid-19 pandemic (another *social disparity* under the pandemic); and 5) People in the Midwest and West regions share similar spending patterns, while people in the Northeast and South regions behave similarly in food spending. From the global perspective, we provide suggestions for government preparation for the pandemic, including international trade and food subcategory associations.

## Broader Impact and Ethical Consideration

Since the pandemic began, there have been different kinds of food shortages over the world. People in different countries display distinct spending behaviors over certain foods, resulting in temporarily lower stocks at grocery stores and higher prices. By analyzing the price changes for different food categories in the United States, we can associate the background of consumers, e.g., location and income, with the shopping patterns under such an epidemic. Apart from investigating the U.S. food price changes, we also mine knowledge from the global food price patterns. Our study provides a better understanding of the relationship between the Covid-19 pandemic and food spending/prices, which can help the Federal and local governments maintain a stable food price level and alleviate the food shortage problem under the fluctuating Covid-19 conditions. Our global analysis sheds light on a better global food supply chain, as the demands of foods could be complementary between different countries. All data collected for the United States, the United Kingdom, China, and Japan are publicly available through the website of their office of national statistics. *Aggregated* data is available upon request in an attempt to promote future work and at the same time minimize the possibility of unethical use and privacy risks.

# References

Baker, S. R.; Farrokhnia, R. A.; Meyer, S.; Pagel, M.; and Yannelis, C. 2020. How does household spending respond to an epidemic? Consumption during the 2020 COVID-19 pandemic. *The Review of Asset Pricing Studies* 10(4): 834– 862.

Chen, H.-S.; Yan, Z.; Yao, Y.; Huang, T.-B.; and Wong, Y.- S. 2018. Systematic procedure for Granger-causality-based root cause diagnosis of chemical process faults. *Industrial & Engineering Chemistry Research* 57(29): 9500–9512.

Deaton, A.; and Paxson, C. 1998. Economies of scale, household size, and the demand for food. *Journal of political economy* 106(5): 897–930.

Deconinck, K.; Avery, E.; and Jackson, L. A. 2020. Food Supply Chains and Covid-19: Impacts and Policy Lessons. *EuroChoices* 19(3): 34–39.

Dickey, D. A.; and Fuller, W. A. 1979. Distribution of the estimators for autoregressive time series with a unit root. *Journal of the American statistical association* 74(366a): 427– 431.

Donnelly, A. 2020. Coronavirus fears: Empty shelves as Canadians heed health minister's advice to stock up.

Dube, L. F.; and Black, G. S. 2010. Impact of national traumatic events on consumer purchasing. *International journal of consumer studies* 34(3): 333–338.

Ellison, B.; McFadden, B.; Rickard, B. J.; and Wilson, N. L. 2021. Examining food purchase behavior and food values during the COVID-19 pandemic. *Applied Economic Perspectives and Policy* 43(1): 58–72.

Ferrante, M. J.; Goldsmith, J.; Tauriello, S.; Epstein, L. H.; Leone, L. A.; and Anzman-Frasca, S. 2021. Food Acquisition and Daily Life for US Families with 4-to 8-Year- Old Children during COVID-19: Findings from a Nationally Representative Survey. *International Journal of Environmental Research and Public Health* 18(4): 1734.

Frank, B.; and Schvaneveldt, S. J. 2014. Self-preservation vs. collective resilience as consumer responses to national disasters: A study on radioactive product contamination. Journal of Contingencies and Crisis Management 22(4): 197–208.

French, S. A.; Wall, M.; and Mitchell, N. R. 2010. House- hold income differences in food sources and food items purchased. *International Journal of Behavioral Nutrition and Physical Activity* 7(1): 1–8.

Glen, S. 2021. Correlation Coefficient: Simple Definition, Formula, Easy Steps.

Granger, C. W. 1969. Investigating causal relations by econometric models and cross-spectral methods. *Econometrica: journal of the Econometric Society* 424–438.

Hobbs, J. E. 2020. Food supply chains during the COVID- 19 pandemic. *Canadian Journal of Agricultural Economics/Revue canadienne d'agroeconomie* 68(2): 171–176.

Hori, M.; and Iwamoto, K. 2014. The run on daily foods and goods after the 2011 Tohoku earthquake: a fact finding analysis based on homescan data. *The Japanese Political Economy* 40(1): 69–113.

Huang, K.-M.; Sant'Anna, A. C.; and Etienne, X. 2021. How did Covid-19 impact US household foods? an analysis six months in. *PloS one* 16(9): e0256921.

Hyndman, R. J.; and Athanasopoulos, G. 2018. *Forecasting: principles and practice*. OTexts.

Kawate, I. 2022. China food prices soar as zero-COVID policy stokes inflation. *Nikkei Asia* .

Knotek II, E. S.; Schoenle, R.; Dietrich, A.; Kuester, K.; Mü̈ller, G.; Myrseth, K. O. R.; and Weber, M. 2020. Consumers and COVID-19: A real-time survey. *Economic Commentary* (2020-08).

Kwiatkowski, D.; Phillips, P. C.; Schmidt, P.; and Shin, Y. 1992. Testing the null hypothesis of stationarity against the alternative of a unit root: How sure are we that economic time series have a unit root? *Journal of econometrics* 54(1- 3): 159–178.

Lusk, J. L. 2019. Income and (Ir) rational food choice. *Journal of Economic Behavior & Organization* 166: 630–645.

McNamara, V. 2021. China's Warnings Spark Panic Buying: The Reasons and Consequences.

Mead, D.; Ransom, K.; Reed, S. B.; and Sager, S. 2020. The impact of the COVID-19 pandemic on food price indexes and data collection. *Monthly Lab*. Rev. 143: 1.

Melo, G. 2020. The Path Forward: US Consumer and Food Retail Responses to COVID-19. *Choices* 35(3).

Oteros-Rozas, E.; Ruiz-Almeida, A.; Aguado, M.; González, J. A.; and Rivera-Ferre, M. G. 2019. A social-ecological analysis of the global agrifood system. *Proceedings of the National Academy of Sciences* 116(52): 26465–26473.

Otsuka, M. 2021. COVID-19 Pandemics Impact on Japan HRI Industry.

Pena-Levano, L.; Burney, S.; Melo, G.; and Escalante, C. 2020. COVID-19: Effects on US Labor, Supply Chains and Consumption Imagery Article. *Choices* 35(3).

Richards, T. J.; and Rickard, B. 2020. COVID-19 impact on fruit and vegetable markets. *Canadian Journal of Agricultural Economics/Revue canadienne d'agroeconomie* 68(2): 189–194.

Roy, S.; and Chakraborty, C. 2021. Panic Buying Situation during COVID-19 Global Pandemic. *Journal of Information Technology Management* 13(2): 231–244.

Statista. 2020. Household grocery expenditure.

Thilmany, D.; Canales, E.; Low, S. A.; and Boys, K. 2021. Local Food Supply Chain Dynamics and Resilience during COVID-19. *Applied Economic Perspectives and Policy* 43(1): 86–104.

U.S. Bureau of Labor Statistics. 2021. Consumer Price Index.

USDA. 2021. Food and Consumers.

Wikipedia. 2020. Census regions and divisions.

Yang, Y.; Peng, F.; Wang, R.; Guan, K.; Jiang, T.; Xu, G.; Sun, J.; and Chang, C. 2020. The deadly coronaviruses: The 2003 SARS pandemic and the 2020 novel coronavirus epidemic in China. *Journal of autoimmunity* 109: 102434.

Ziliak, J. P. 2021. Food hardship during the COVID-19 pandemic and Great Recession. *Applied Economic Perspectives and Policy* 43(1): 132–152.